\documentclass{iopart}
\usepackage{graphicx}
\usepackage{iopams}

\begin{document}
\title[Strongly coupled QGP at RHIC]{The strongly coupled quark-gluon 
       plasma created at RHIC\footnote[7]{Work supported by U.S. 
       Department of Energy, grant DE-FG02-01ER41190.}
}
\author{Ulrich Heinz$^{1,2}$}
\address{$^1$Department of Physics, The Ohio State University, Columbus, 
         OH 43210, USA}
\address{$^2$CERN, Physics Department, Theory Division,
         CH-1211 Geneva 23, Switzerland}


\begin{abstract}
The Relativistic Heavy Ion Collider (RHIC) was built to re-create and 
study in the laboratory the extremely hot and dense matter that filled 
our entire universe during its first few microseconds. Its operation 
since June 2000 has been extremely successful, and the four large RHIC 
experiments have produced an impressive body of data which indeed provide 
compelling evidence for the formation of thermally equilibrated matter at
unprecedented temperatures and energy densities -- a "quark-gluon plasma 
(QGP)". A surprise has been the discovery that this plasma behaves like 
an almost perfect fluid, with extremely low viscosity. Theorists had 
expected a weakly interacting gas of quarks and gluons, but instead we 
seem to have created a strongly coupled plasma liquid. The experimental 
evidence strongly relies on a feature called "elliptic flow" in 
off-central collisions, with additional support from other observations. 
This article explains how we probe the strongly coupled QGP, describes 
the ideas and measurements which led to the conclusion that the QGP is 
an almost perfect liquid, and shows how they tie relativistic heavy-ion 
physics into other burgeoning fields of modern physics, such as strongly 
coupled Coulomb plasmas, ultracold systems of trapped atoms, and 
superstring theory.
\end{abstract}

\section{Introduction: what is a QGP, and how to create it}
\label{sec1}

Initially, the Universe didn't look at all like it does today: until about
10\,$\mu$s after the Big Bang, it was too hot and dense to allow quarks
and gluons to form hadrons, and the entire Universe was filled with a 
thermalized plasma of deconfined quarks, antiquarks, gluons and leptons 
(and other, heavier particles at even earlier times) -- a Quark-Gluon 
Plasma (QGP). After hadronization of this QGP, it took another 3 minutes 
until the first small nuclei were formed from protons and neutrons 
(primordial nucleosynthesis and chemical freeze-out), another 400,000 
years until atomic nuclei and electrons could combine to form electrically 
neutral atoms, thereby making the Universe transparent and liberating the 
Cosmic Microwave Background (kinetic decoupling of photons and thermal 
freeze-out), and another 13 billion years or so for creatures to 
evolve with sufficient intelligence to contemplate all of this.

With high-energy nuclear collisions at the Relativistic Heavy Ion Collider
(RHIC) at Brookhaven National Laboratory or at the Large Hadron Collider 
(LHC) at CERN, one attempts to recreate the conditions of the Early Universe
in the laboratory, by heating nuclear matter back up to temperatures
above the quark-gluon deconfinement temperature. The hot and dense 
fireballs created in these collisions undergo thermalization, cooling by 
expansion, hadronization, chemical and thermal freeze-out in a pattern
that strongly resembles the evolution of the Universe after the Big Bang.
For this reason heavy-ion collisions are often called ``Little Bangs''.
A major difference is, however, the much smaller size of the
Little Bang and its much (by about 18 orders of magnitude) faster 
dynamical evolution when compared with the Big Bang. This complicates
the theoretical analysis of the experimental observations.   

The main stages of the Little Bang are: 1.) Two disk-like nuclei approach 
each other, Lorentz contracted along the beam direction by a factor 
$\gamma=E_\mathrm{beam}/M$ (where $E_\mathrm{beam}$ is the beam energy 
per nucleon and $M=0.94$\,GeV is the nucleon mass, such that $\gamma\approx
110$ for heavy ions at RHIC and $\approx 3000$ at the LHC at their 
respective top energies). 2.) After impact, hard collisions which large 
momentum transfer $Q \gg 1$\,GeV between quarks, antiquarks or gluons 
(partons) inside the nucleons of the two nuclei produce secondary partons 
with large transverse momenta $p_T$ at early times $\sim 1/Q \sim 1/p_T$. 
3.) Soft collisions with small momentum exchange $Q\lesssim 1$\,GeV 
produce many more particles somewhat later and thermalize the QGP after 
about 1\,fm/$c$. The resulting thermalized QGP fluid expands 
hydrodynamically and cools approximately adiabatically. 4.) The QGP 
converts to a gas of hadrons; the hadrons continue to interact 
quasi-elastically, further accelerating the expansion and cooling 
of the fireball until thermal freeze-out. The chemical composition of 
the hadron gas is fixed during the hadronization process and remains
basically unchanged afterwards. After thermal decoupling, unstable 
hadrons decay and the stable decay products stream freely towards the 
detector.

By studying the behaviour of the matter created in the Little Bangs we 
can explore the phase structure and phase diagram of strongly 
interacting matter. Where the proton-proton collision program at the
LHC aims at an understanding of the elementary degrees of freedom
and fundamental forces at the shortest distances, the heavy-ion program
focusses on the condensed matter aspects of bulk material whose constituents
interact with each other through these forces. The difference between
this kind of condensed matter physics and the traditional one is that in
our case the fundamental interaction is mediated by the strong rather than 
the electromagnetic force. The coupling strength of the strong interaction 
gets bigger rather than smaller at large distances, leading us to expect
a completely new type of phase structure. Indeed, strongly interacting
matter appears to behave like a liquid (``quark soup'') at high temperature 
and like a gas (``hadron resonance gas'') at low temperature, contrary to 
intuition. On the other hand, the QGP state, with its unconfined color 
charges, has similarities with electrodynamic plasmas whose dynamical
behaviour is controlled by the presence of unconfined electric charges.
For example, both feature Debye screening of (color) electric fields. 
The main differences are QGP temperatures that are about a factor 1000 
higher, and particle densities that are about a factor $10^9$ larger,
than their counterparts in the hottest and densest electrodynamic plasmas.
Accordingly, quark-gluon plasmas are intrinsically relativistic, and 
magnetic fields play everywhere an equally important role as electric
fields. The nonlinear effects resulting from the non-Abelian structure
of the strong interaction manifest themselves most importantly in the
magnetic sector.
 
The equilibrium properties of a QGP with small net baryon density can 
be computed from first principles using Lattice QCD. Such calculations
exhibit a smooth but rapid cross-over phase transition from a QGP
above $T_c$ to a hadron gas below $T_c$ around critical temperature
$T_c= 173\pm15$\,MeV, corresponding to a critical energy density
$e_c\simeq 0.7$\,GeV/fm$^3$. For $T\gtrsim2\,T_c$, the QGP
features an equation of state $p(e)$ corresponding to an ideal gas
of massless partons, $p\approx\frac{1}{3}e$, with sound speed $c_s=
\sqrt{\frac{\partial p}{\partial e}}=\sqrt{\frac{1}{3}}$. But both the 
normalized energy density $e/T^4$ and pressure $p/T^4$ remain about
15-20\% below the corresponding Boltzmann limits for a non-interacting
massless parton gas. At face value, this relatively small deviation from 
the ideal gas appears to support the idea (long held by theorists before
the RHIC era) that, because of ``asymptotic freedom'' (the QCD coupling 
``constant'' decreases logarithmically with increasing energy), the QGP 
is a weakly interacting system of quarks, antiquarks and gluons that can 
be treated perturbatively. However, RHIC experiments proved this 
expectation to be quite wrong. Instead the QGP was found to behave like 
an almost perfect liquid, requiring it to be a {\em strongly coupled 
plasma}.   

This experimental surprise is based on three key observations: 1.) 
{\bf Large elliptic flow:} The anisotropic collective flow of the 
fireball matter measured in non-central heavy-ion collisions is huge 
and essentially exhausts the upper theoretical limit predicted by 
ideal fluid dynamics. 2.) {\bf Heavy quark collective flow:} Even 
heavy charm and bottom quarks are observed to be dragged along by 
the collective expansion of the fireball and exhibit large elliptic 
flow. 3.) {\bf Jet quenching:} Fast partons plowing through the 
dense fireball matter, even if they are very heavy such as charm and 
bottom quarks, lose large amounts of energy, leading to a strong 
suppression of hadrons with large $p_T$ when compared with expectations 
based on a naive extrapolation of proton-proton collision data, and 
the quenching of hadronic jets.

As I will try to explain in the rest of this review, none of these 
three observations can be understood without assuming some sort of 
strong coupling among the plasma constituents. Further, I will
show that there is rather compelling evidence that we are indeed 
dealing with a ``plasma'' state where color charges are deconfined. 
The strongly-coupled nature of the QGP manifests itself in its 
collectivity and its transport properties, seen in non-equilibrium
situations such as those generated in heavy-ion collisions. It is much
more difficult to extract directly from its bulk equilibrium properties
which are studied in lattice QCD.

Statements made and conclusions drawn in this overview are based
on a large body of experimental data and comparisons with models for 
the dynamical evolution of heavy-ion collisions. Unfortunately,
the assigned space does not permit me to show these here. For the most
important plots and a comprehensive list of references I refer the 
interested reader to recent reviews by M\"uller and Nagle 
\cite{Muller:2006ee} and by Shuryak \cite{Shuryak:2008eq}. For some 
newer developments I will provide references to the original papers. 

\section{Collective flow -- the ``Bang''}
\label{sec2}

The primary observables in heavy-ion collisions are (i) hadron momentum 
spectra and yield ratios, from which we can extract the chemical 
composition, temperature, and collective flow pattern (including 
anisotropies) of the fireball when it finally decouples, (ii) the energy
distributions of hard direct probes such as jets created by hadronizing
fast partons, hadrons containing heavy charm and bottom quarks, and 
directly emitted electromagnetic radiation, from which we can extract 
information on the fireball temperature and density at earlier times 
before it decouples, and (iii) two-particle momentum correlations from 
which we can extract space-time information about the size and shape of 
the fireball at decoupling (hadron correlations) or at earlier times 
(photon correlations). Except for photon correlations which are hard
to measure, good experimental data exist now for all of these 
observables. To learn about the existence and properties of the 
quark-hadron phase transition itself one needs to study event-by-event
fluctuations around the statistical average of the event ensemble; 
this subject is still in its infancy, both experimentally and 
theoretically.

I will first discuss the collective flow patterns observed in 
heavy-ion collisions at RHIC. Collective flow is driven by pressure
gradients and thus provides access to the equation of state $p(e)$
($p$ is the thermodynamic pressure, $e$ is the energy density, and the 
net baryon density $n_B$ is small enough at RHIC energies that its 
influence on the EOS can be neglected). The key equation is
\begin{equation}
{\dot u}^\mu=\frac{\nabla^\mu p}{e+p} = \frac{c_s^2}{1+c_s^2}
             \frac{\nabla^\mu e}{e},
\end{equation}
valid for ideal fluids, which shows that the acceleration of the fluid
is controlled by the speed of sound $c_s=\sqrt{\frac{\partial p}{\partial e}}$
(reflecting the stiffness of the EOS $p(e)$) which determines the fluid's
reaction to the normalized pressure or energy density gradients. (The dot
in ${\dot u}^\mu \equiv (u\cdot\partial)u^\mu$ denotes the time derivative 
in the fluid's local rest frame (LRF), and $\nabla^\mu$ is the spatial 
gradient in that frame. For non-ideal (viscous) fluids, additional terms 
appear on the r.h.s., depending on spatial velocity gradients in the LRF
multiplied by transport coefficients (shear and bulk viscosity).) 

Lattice QCD data show that $c_s^2$ decreases from about $\frac{1}{3}$ at 
$T>2T_c$ by almost a factor 10 close to $T_c$, rising again to around 
$0.16-0.2$ in the hadron gas (HG) phase below $T_c$. The finally 
observed collective flow transverse to the beam direction reflects
a (weighted) average of the history of $c_s^2(T)$ along the cooling
trajectory explored by the fireball medium. Different aspects of the
final flow pattern weight this history differently. Whereas the 
azimuthally averaged ``radial flow'' receives contributions from all
expansion stages, due to persistent (normalized) pressure gradients 
between the fireball interior and the outside vacuum, flow anisotropies,
in particular the strong ``elliptic flow'' seen in non-central collisions,
are generated mostly during the hot early collision stages. They are
driven by spatial anisotropies of the pressure gradients due to the
initial spatial deformation of the nuclear reaction zone (see Fig.~1), 
%
\begin{figure}[hb]
  \begin{center}
  \includegraphics[width=0.7\linewidth,clip=]
                  {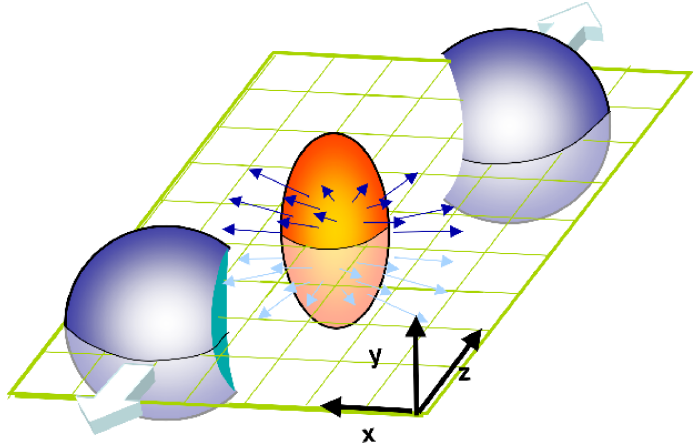}
  \end{center}
   \caption{\label{spectra}(Color online)
   Sketch of deformed fireball and elliptic flow generated in
   non-central heavy-ion collisions.}
\end{figure}
%
but this 
deformation decreases with time as a result of anisotropic flow, since 
the matter accelerates more rapidly, due to larger pressure gradients, 
in the direction where the fireball was initially shorter. With the
disappearance of pressure gradient anisotropies the driving force for
flow anisotropies vanishes, and due to this ``self-quenching'' effect
the elliptic flow saturates early. If the fireball expansion starts
at sufficiently high initial temperature, it is possible that all
elliptic flow is generated before the matter reaches $T_c$ and hadronizes.
In this case (which we expect to be realized in heavy-ion collisions at
the LHC) elliptic flow is a clean probe of the EOS of the QGP phase.

Elliptic flow is measured as the second Fourier coefficient of the
azimuthal angle distribution of the final hadrons in the transverse
plane:
\begin{equation}  
v_2(y,p_T;b) = \langle \cos(2\phi_p)\rangle
= \frac{\int d\phi_p \cos(2\phi_p) \frac{dN}{dy\,p_T\,dp_T\,d\phi_p}(b)}
       {\int d\phi_p               \frac{dN}{dy\,p_T\,dp_T\,d\phi_p}(b)}.
\end{equation}
Here $y=\frac{1}{2}\ln[(E+p_L)/(E-p_L)]$ is the rapidity of the 
particles, and $b$ the impact parameter of the collision. Each particle
species has its own elliptic flow coefficient, characterizing the elliptic
azimuthal deformation of its momentum distribution. $v_2$ has been measured
as a function of transverse momentum $p_T$ for a variety of hadron species
with different masses, ranging from the pion ($m_\pi=140$\,MeV) to
the $\Omega$ hyperon ($m_\Omega=1672$\,MeV). For some hadron species 
$v_2$ has been measured out to $p_T>10$\,GeV where the spectrum has 
decayed by more than 7 orders of magnitude from the yield
measured at low $p_T$! More than 99\% of all hadrons have $p_T<2$\,GeV;
in this domain the data show excellent agreement with ideal fluid dynamical 
predictions, including the hydrodynamically predicted rest mass dependence 
of $v_2$ (at the same $p_T$, heavier hadrons show less elliptic flow). 
Ideal fluid dynamics thus gives a good description of the collective 
behaviour of the bulk of the fireball matter.  

It must be emphasized that the ideal hydrodynamic prediction of $v_2(p_T)$
is essentially parameter free: All model parameters (initial conditions
and decoupling temperature) are fixed in central collisions where $v_2=0$, 
and the only non-trivial input for non-central collisions is the initial
geometric source eccentricity as a function of impact parameter. Originally,
one computed this eccentricity from a geometric Glauber model, and in
this case one finds that the experimental data fully exhaust the 
theoretical prediction from ideal fluid dynamics, leaving
very little room for viscosity which would reduce the theoretical value
for the elliptic flow. This is the cornerstone of the ``perfect fluid''
paradigm for the QGP that has emerged from the RHIC data. However, recently 
suggested alternate models for the initial state, for example the Color Glass 
Condensate (CGC) model, can give initial eccentricities that are up to 30\% 
larger than the Glauber model values. Furthermore, the first ideal fluid 
calculations used an incorrect chemical composition during the late 
hadronic stage of the collision; once corrected, this increased the 
theoretical prediction for $v_2(p_T)$ for pions by another 30\% or so. 
If both of these effects are included, the measured $v_2(p_T)$ reaches 
only about 2/3 of the ideal fluid limit, opening some room for viscous 
effects in the fireball fluid. 

At RHIC energies, not all of the flow anisotropy is created before 
hadronization, especially in noncentral collisions and away from 
midrapidity where the initial temperatures are not as high as in central 
collisions near midrapidity. Recent studies which treat this late hadronic
stage microscopically rather than as an ideal fluid have exhibited large
viscous effects in the hadron gas phase \cite{Hirano:2005xf}. If these 
are taken into account in the theoretical description, one finds that 
they reduce the elliptic flow and compensate for the $\sim 30\%$ increase 
of $v_2$ due to non-equilibrium hadron chemistry mentioned above. With 
Glauber initial conditions one is thus again left with almost no room 
for QGP viscosity, whereas CGC initial conditions allow the QGP fluid 
to have some non-zero viscosity.\footnote{An important aspect of the 
hydrodynamical simulations that describe the experimental data well is 
that they require short thermalization times, of order 1\,fm/$c$.
This is true in particular for Glauber initial conditions where one
simply doesn't get enough elliptic flow to describe the data if one
doesn't initiate the hydrodynamic expansion before 1\,fm/$c$ (where the 
clock starts at nuclear impact). Short thermalization times and the 
validity of an ideal fluid picture are, of course, two sides of the 
same consistent picture. For CGC initial conditions, assuming similarly 
short thermalization times, the experimental $v_2$ data do not saturate 
the ideal fluid prediction. In this case one can either start the ideal 
fluid dynamic expansion later, or endow the fluid with some non-zero 
viscosity, or both. Again, these are two sides of the same consistent 
picture, which now invokes non-zero mean free paths for the plasma 
constituents, leading to incomplete local thermalization.}

Recent theoretical progress in the formulation and simulation of 
relativistic hydrodynamics for {\em viscous} fluids (see 
\cite{Heinz:2008qm} for a summary) has provided us 
with a tool that permits us to answer the question how large the QGP
viscosity might be. This breakthrough is based on second-order
Israel-Stewart theory and variations thereof which avoids longstanding
problems of violations of causality in relativistic Navier-Stokes theory
which includes only first-order gradients of the local thermodynamic
variables. This is still largely work in progress, and published results
based on this approach do not yet include all the physical ingredients
known to be relevant for a quantitative prediction of elliptic flow. 
Nonetheless, a first heroic effort has been made by Luzum and Romatschke
\cite{Luzum:2008cw} to use this new approach to limit the shear viscosity 
to entropy ratio $\eta/s$ from experimental elliptic flow data. Their work 
does not include non-equilibrium chemistry in the late hadronic stage 
(which would increase the calculated $v_2$), nor does it subtract effects 
from increased viscosity during that stage (which would reduce $v_2$). 
These two effects work against each other, and it may therefore not be 
too presumptuous to try to extract an upper limit for $\eta/s$ from their
results, shown in Fig. 8 of Ref.~\cite{Luzum:2008cw}. Even with all the 
uncertainties shown in that Figure (Glauber vs. CGC and a claimed 20\%
uncertainty in the normalization of the experimental data), it looks
like $\frac{\eta}{s}>3 \left(\frac{\eta}{s}\right)_\mathrm{min}$
(where $\left(\frac{\eta}{s}\right)_\mathrm{min}=\frac{1}{4\pi}\approx
0.08$ is a conjectured absolute lower limit on the specific shear
viscosity for {\em any} fluid, derived for strongly coupled conformally
invariant supersymmetric field theories using the AdS/CFT correspondence
\cite{Kovtun:2004de}) is difficult to accomodate. (The authors quote a 
conservative upper limit of $\eta/s<0.5$.) Since all known classical 
fluids have $\eta/s >> 1$ at all temperatures, even at their boiling 
points where $\eta/s$ typically reaches a minimum \cite{Kovtun:2004de}, 
this makes the QGP the most perfect fluid of any studied so far in the 
laboratory. (Recent studies indicate that ultracold atoms in the unitary 
limit (infinite scattering length) may come in a close second
\cite{Rupak:2007vp}.) 
  
\section{Primordial hadrosynthesis -- measuring $T_c$}
\label{sec3}
 
The observed hadron yields (better: hadron yield ratios, from which the
hard to measure fireball volume drops out) tell us about the chemical
composition of the fireball when it finally decouples. It turns out that 
the hadron yield ratios measured in Au+Au collisions at RHIC can be 
described extremely well using a thermal model with just two parameters: 
a chemical decoupling temperature $T_\mathrm{chem}=163\pm4$\,MeV
and a small baryon chemical $\mu_B=24\pm4$\,MeV. In central and semi-central
collisions the phase space for strange quarks is fully saturated -- if
one generalizes the thermal fit to include a strangeness saturation factor 
$\gamma_s$ one finds $\gamma_s=0.99\pm0.07$. In peripheral collisions
with less than 150 participating nucleons, $\gamma_s$ is found to drop,
approaching a value around 0.5 in $pp$ collisions, reflecting the well-known
strangeness suppression in such collisions. This suppression is completely
removed in central Au+Au collisions. In contrast to $\gamma_s$, the
chemical decoupling temperature $T_\mathrm{chem}$ is found to be 
completely independent of collision centrality. So, at freeze-out, all 
Au+Au collisions are well described by a thermalized hadron resonance
gas in {\em relative} chemical equilibrium with respect to all types
of inelastic, identity-changing hadronic reactions, as long as the total
number of strange valence quark-antiquark pairs (which in peripheral 
collisions is suppressed below its {\em absolute} equilibrium value) 
is conserved.

Two aspects of this observation are, at first sight, puzzling: (i) The
measured chemical decoupling temperature is, within errors, consistent
with the critical temperature for hadronization of a QGP predicted by 
lattice QCD. If hadrons formed at $T_c$ and then stopped interacting
inelastically with each other at $T_\mathrm{chem}\approx T_c$, how was 
there ever enough time in the continuously expanding and cooling fireball 
for their abundances to reach chemical equilibrium? (ii) If chemical
equilibrium among the hadrons is controlled by inelastic scattering 
between them, the decoupling of hadron abundances is controlled by a
competition between the microscopic inelastic scattering rate and the
macroscopic hydrodynamic expansion rate. Since the expansion rate depends
on the fireball radius and thereby on the impact parameter of the Au+Au
collision, the chemical decoupling temperature (which controls the
density and thus the scattering rate) should also depend on collision
centrality \cite{Heinz:2007in}. How can the measured $T_\mathrm{chem}$ 
then be independent of centrality?

There is only one explanation that resolves both puzzles simultaneously:
The hadrons are born into a maximum entropy state by complicated 
quark-gluon dynamics during hadronization, and after completion
of this process the hadronic phase is so dilute that inelastic hadronic 
collisions (other than resonance scattering that only affects the particles'\
momenta but not the abundances of finally observed stable decay products) 
can no longer compete with dilution by expansion, freezing the chemical 
composition. This maximum entropy state cannot be distinguished from the 
chemical equilibrium state with the same temperature and chemical potential
that would eventually be reached by microscopic inelastic scattering if
the fireball medium were held in box. The measured temperature 
$T_\mathrm{chem}=T_c$ is, however, not established by hadronic scattering 
processes, but characterizes the energy density at which the quarks and 
gluons coalesce into hadrons, independent of the local expansion rate of
the fluid in which this happens. The microscopic dynamics itself that 
leads to the observed maximum entropy state is not describable in terms 
of well-defined hadronic degrees of freedom but involves effective 
degrees of freedom which control the microscopic physics of the 
hadronization phase transition. 

The absence of inelastic hadronic scattering processes after completion 
of hadronization is fortunate since it opens a window onto the phase 
transition itself, allowing us to measure $T_c$ through the 
hadron yields in spite of the fact that the hadrons continue to scatter
quasi-elastically, maintaining approximate thermal (but not chemical!) 
equilibrium for the momentum distributions down to much lower thermal 
decoupling temperatures around 100\,MeV. This ``kinetic decoupling 
temperature'' can be extracted from the measured momentum distributions
and {\em is} found to depend on collision centrality, as predicted by
hydrodynamics \cite{Heinz:2007in}.
  

\section{JET: Jet Emission Tomography of the QGP}
\label{sec5}

As explained so far, RHIC collisions show strong evidence for fast 
thermalization of the momenta of the fireball constituents throughout 
at least the earlier part of the fireball expansion (until hadronization) 
and for chemical equilibrium at hadronization. After hadronization, 
chemical equilibrium is immediately broken at $T_c$, and thermal 
equilibration becomes gradually less efficient until the momenta finally
decouple, too, a $T_\mathrm{therm}\sim 100\,\mathrm{MeV} < T_\mathrm{chem}$. 

What causes the fast thermalization during the early expansion stage?
We can use fast partons, created in primary collisions between quarks 
or gluons from the two nuclei, to probe the early dense stage of the
medium. Such hard partons, emitted with high transverse momenta $p_T$,
fragment into a spray of hadrons in the direction of the parton, forming
what's called a {\em jet}. The rate for creating such jets can 
be factored into a hard parton-parton cross section, described by 
perturbative QCD, a soft structure function describing the probability
to find a parton to scatter off inside a nucleon within the colliding 
nuclei, and a soft fragmentation function describing the fragmentation 
of the scattered parton into hadrons. The structure and fragmentation 
functions are universal and can be measured in deep-inelastic $ep$
scattering (DIS) and in $pp$ collisions.
Nuclear modifications of the structure function can be measured in
DIS of electrons on nuclei. Jets thus form a calibrated, selfgenerated 
probe which can used to explore the fireball medium tomographically.
The medium will affect the hard parton along the path from its production
point to where it exits the fireball. If the parton is sufficiently
energetic, it will exit the medium before it can begin to fragment into 
hadrons. The difference in jet production rates or, more generally,
in the rates for producing high-$p_T$ hadrons from jet fragmentation 
in Au+Au and $pp$ collisions can be calculated in terms of the 
density of scatterers in the medium, multiplied with a perturbatively 
calculable cross section (if the parton has sufficiently high energy
to justify a perturbative approach), and integrated along the path of the
jet. This integrated product of density times cross section characterizes
the opacity of the medium. Since the probe is colored and interacts
through color exchange, it is sensitive to density of color charges
resolvable at the scale of its Compton wave length; that density will be 
higher in a color-deconfined QGP than in a cold nucleus where quarks and 
gluons are hidden away inside the nucleons. 

The procedure is similar to the familiar Positron Emission Tomography 
(PET) in medicine -- therefore I call it ``JET''. The main difference 
is that the positron emitting source used in PET is injected externally 
into the medium to be probed while the jets used in heavy-ion collisions 
are created internally together with the medium.

One of the first results from RHIC, right after the discovery of strong
elliptic flow, was the experimental confirmation of the theoretical
prediction that QGP formation should result in a strong suppression
of high-$p_T$ hadrons compared to appropriately scaled $pp$ collisions. 
The observed suppression amounts to a factor 5-6, almost independent of
$p_T$ in the region $4<p_T<15$\,GeV/$c$. This is close to $A^{1/3}$ for 
$A\sim 200$ and suggests a surface/volume effect, i.e. that high-$p_T$ 
hadrons are predominantly emitted from the firball surface while fast 
partons created in the fireball interior lose so much energy before
exiting that they no longer contribute to high-$p_T$ hadron production.
Indeed, angular correlations between a high-$p_T$ leading hadron
selected from the collision and other energetic hadrons, indicative
of a fragmenting jet, strongly support such a picture: Whereas in
$pp$ and d+Au collisions, where fast sideward-moving partons escape
from the narrow ``fireball medium'' almost instantaneously (if one can 
even use the notion of a ``medium'' in that case), one observes two such
correlation peaks, separated by $180^\circ$ and corresponding to the
pair of hard partons created back-to-back in the primary hard scattering,
one sees only one such peak in central Au+Au collisions, in the direction 
of the fast trigger haddron. This can be understood if one assumes that
the trigger hadron stems from the fragmentation of the outgoing partner
of a hard parton pair created near the surface of the fireball, which
exits the medium soon after creation, while its inward-travelling 
partner at $180^\circ$ loses most of its energy before exiting the 
fireball on the other side and no longer contributes energetic hadrons
correlated with the trigger hadron. Still, the energy initially carried
by that partner should show up near $180^\circ$ relative to the trigger
hadron, and it does: While the away-side correlations of energetic 
hadrons with the trigger one are {\em depleted}, the away-side 
correlations between ``soft'' hadrons ($p_T<1.5$\,GeV/$c$) and the 
trigger hadron are {\em enhanced}. The energy lost by the fast parton
travelling away from the trigger hadron and through the medium re-appears
in the form of additional soft hadrons, with a distribution of transverse 
momenta similar to that of the medium itself: As the Au+Au collisions
become more central, the average transverse momenta $\langle p_T\rangle$
of the extra hadrons emitted into the away-side hemisphere are observed
to approach the $\langle p_T\rangle$ of the entire collision event.

In non-central collisions, the fireball created in the collision is 
deformed like an egg. Its orientation can be determined
from the elliptic flow of the soft hadrons discussed in Sec.~\ref{sec2}.
In this case, the overall fireball size is smaller than in central Au+Au
collisions, and the observed away-side suppression of jet-like angular 
correlations is less complete. (Indeed, even in central Au+Au collisions,
the away-side angular correlations among hard hadrons begin to reappear
when the energy of the trigger hadron is increased beyond 10\,GeV or so;
this is apparently too high for the inward-traveling partner to lose all
of its initial energy.) But the suppression that is observed is stronger
if the trigger is emitted perpendicular to the reaction plane (and its 
partner thus must travel through the fire-egg along its long direction)
than when it moves within the reaction plane (i.e. its partner passes
through the fire-egg along its short direction). 

Fast partons moving through a QGP collide with its constituents, causing 
them to lose energy via both elastic collisions and collision-induced 
gluon radiation. For very energetic partons radiative energy loss is 
expected to dominate, so first model comparisons with the data included
only radiative energy loss. From such calculations it was concluded that
the observed suppression of hard particles required densities of scatterers
that were consistent with and independently confirmed estimates of
the initial energy densities extracted from the successful hydrodynamical
models that describe the measured elliptic flow of soft hadrons. On a
more quantitative level, radiative energy loss models soon started, however,
to develop difficulties. They could not reproduce the observed large 
difference in away-side jet quenching between the in-plane and 
out-of-plane emission directions. Decay electrons from weak decays of
hadrons containing charm and bottom quarks and pointing back to these
hadrons were observed to feature strong elliptic flow (indicating
that even these heavy quarks (``boulders in the stream'') are dragged 
along by the anisotropically expanding QGP liquid) and large energy 
loss, almost as large as that observed for hadrons containing only light 
quarks. The inclusion of elastic collisional energy loss and recent 
refinements in the calculation of radiative energy loss have reduced
the disagreement between theory and experiment, but some significant
tension remains. This has recently generated lively theoretical activity 
aiming to compute heavy quark drag and diffusion coefficients for strongly 
coupled quantum field theories, using gravity duals and the AdS/CFT 
correspondence (see \cite{Shuryak:2008eq} for a review and references). 

What happens to all the energy lost by fast partons plowing through
a QGP? There is some experimental evidence (and it appears to be getting 
stronger with recent 3-particle correlation measurements) that the soft 
partons emitted into the away-side hemisphere relative to a hard trigger 
particle emerge in the shape of a conical structure. This could signal
a Mach shock cone, generated by the supersonically moving fast parton as
it barrels through the perfect QGP liquid. Interestingly, the perhaps most
convincing theoretical approach that actually generates something like
the observed structures in the angular and 3-particle correlations on
the away-side of the trigger jet again is based on models using gravity
duals and the AdS/CFT connection \cite{Chesler:2007sv}. Clearly this 
needs more work, but the implications are intriguing. 

\section{Outlook}
\label{sec6}

On a qualitative level, the new RHIC paradigm, which states that the QGP
is a strongly coupled plasma exhibiting almost perfect liquid behaviour 
and strong color opacity (even for the heaviest colored probes such as
charm and bottom quarks), has solid experimental and theoretical support.
The microscopic origins of the strong coupling observed in the collective
dynamical behaviour of the QGP remain, however, to be clarified. Theorists 
are approaching this question from three angles: perturbative QCD based on
an expansion in $\alpha_s\ll 1$, lattice QCD with $\alpha_s$ adjusted
to reproduce the measured hadron mass spectrum, and strong-coupling 
methods for the limit $\alpha_s\gg 1$, based on the AdS/CFT correspondence
which states that properties of certain strongly coupled field theories 
can be calculated by solving Einstein's equations in appropriately curved
space-times called ``gravity duals''. Quantitatively precise
and reliable results from either approach are expected to still require 
much hard work. An overview over some of the ongoing theoretical activities
in this direction can be found in \cite{Shuryak:2008eq}.

However, even without a complete quantitative theoretical understanding
of the QGP transport coefficients, the body of experimental heavy-ion 
collision data is already very rich, and it is expected to further grow at 
a staggering rate with the completion of the RHIC II upgrade and the turn-on
of the LHC. With the continued development of increasingly sophisticated 
models for the fireball expansion dynamics and its ultimate decoupling 
into non-interacting hadrons, the time is ripe for a comprehensive attack
on the problem of extracting precise values for the QGP transport
coefficients from a phenomenological description of the experimental
data. This program has already started and produced first preliminary
results as reported here; its outcome will yield valuable constraints 
and guidance for the theorists aiming for a first-principles 
based theoretical understanding of the QGP and its properties.

{\bf Acknowledgement:} This work was supported by the U.S. Department 
of Energy under Contract DE-FG02-01ER41190. I am grateful to my students
and postdocs who joined me in this research for so many years and helped
develop a coherent picture of the complex dynamics of heavy-ion collisions.  


\end{document}